# Double Stage Delay Multiply and Sum Beamforming Algorithm Applied to Ultrasound Medical Imaging

Moein Mozaffarzadeh[a], Masume Sadeghi[a], Ali Mahloojifar[a,*], Mahdi Orooji[a]

[a]*Department of Biomedical Engineering, Tarbiat Modares University, Tehran, Iran*

**Abstract**

In Ultrasound (US) imaging, Delay and Sum (DAS) is the most common beamformer, but it leads to low quality images. Delay Multiply and Sum (DMAS) was introduced to address this problem. However, the reconstructed images using DMAS still suffer from level of sidelobes and low noise suppression. In this paper, a novel beamforming algorithm is introduced based on the expansion of DMAS formula. It is shown that there is a DAS algebra inside the expansion, and it is proposed to use DMAS instead of the DAS algebra. The introduced method, namely Double Stage DMAS (DS-DMAS), is evaluated numerically and experimentally. The quantitative results indicate that DS-DMAS results in about 25% lower level of sidelobes compared to DMAS. Moreover, the introduced method leads to 23%, 22% and 43% improvement in Signal-to-Noise Ratio, Full-Width-Half-Maximum and Contrast Ratio, respectively, in comparison with DMAS beamformer.

---

**Introduction**

Ultrasound (US) imaging is one of the most common medical imaging modalities due to its low cost and high safety (Hansen et al., 2014). The quality of US images highly depends on the beam properties. The presence of off-axis signals in the image reconstruction procedure leads to resolution and contrast degradation in the reconstructed images (Ranganathan and Walker, 2003). In US imaging, Delay and Sum (DAS) is a common beamforming algorithm due to its simple implementation and capability of real time imaging. However, it results in low resolution images, high level of sidelobes and limited off-axis signal rejection (Karaman et al., 1995). A variety of methods and algorithms are used to improve the performance of beamforming in US imaging. One of the alternatives for DAS is Minimum Variance (MV) (Asl and Mahloojifar, 2009). Even though MV provides a higher resolution in comparison with DAS, it imposes a high computational burden. Moreover, MV leads to high level of sidelobes which degrades the quality of the reconstructed images. MV adaptive beamformer has been modified over the past few years in different aspects such as computational burden reduction (Deylami and Asl, 2016; Bae et al., 2016). Phase screen aberration correction has been combined with MV beamforming algorithm in order to compensate the effects of the sound-velocity heterogeneity which is a significant issue in medical US imaging (Ziksari and Asl, 2017). Another method for image quality enhancement is apodization which is a common way to control the sidelobes of the beam pattern in cases where the imaging medium is considered to be homogeneous, and sound velocity is assumed constant. However, in practice, the sound velocity may vary for about 5%, which highly affects the beam-forming performance. An optimization problem robust to the speed variations was used to minimize the sidelobe levels while maintaining the width of mainlobe (Gholampour et al., 2016). Pixel-based focusing technique has been used in the US linear-array imaging based on restricting the number of different sub-aperture positions. This method selects the best-possible signal for data superposition, and it leads to a higher image quality and resolution in comparison with DAS (Nguyen and Prager, 2016). The main problem with DAS algorithm is its blindness which can be addressed by extension of the receive aperture length in phased synthetic aperture imaging (Sadeghi and Mahloojifar, 2017). Recently, Delay and Standard Deviation (DASD) beamforming algorithm was introduced in order to address the relatively poor appearance of the interventional devices such as needles, guide wires, and catheters in the conventional US images (Bandaru et al., 2016). Three-dimensional US imaging is an emerging technique with a high accuracy in diagnosis. However, it imposes a high computational complexity. More recently, a method was proposed based on decomposing the delay term in a way that it minimizes the root-mean-square error caused by the decomposition. It helps reduction of computational burden of Three-dimensional US imaging (Yang et al., 2015). Beamforming in the Fourier domain can be used as a method for reducing the computational complexity, and achieving fast and accurate image reconstruction (Kruizinga et al., 2012).

In 2015, Delay-Multiply-and-Sum (DMAS) was introduced for US imaging by Matrone *et al.* (Matrone et al., 2015). This algorithm was initially used as a reconstruction algorithm in confocal microwave imaging for breast cancer detection (Lim et al., 2008). DMAS has been used with Multi-Line Transmis-sion (MLT) for high frame-rate US imaging (Matrone et al., 2016). The loss of brightness in DMAS was compensated using a particular implementation of synthetic aperture focusing technique, namely Synthetic Transmit Aperture (STA), combined with DMAS. The main enhancement gained by DMAS is the higher contrast and lower sidelobes. Although it leads to a higher resolution compared to DAS, the resolution is not good enough in comparison with the resolution gained by MV-based algorithms. Thus, MV beamformer has been combined with DMAS algorithm to improve the resolution of DMAS (Mozaffarzadeh et al., 2017d,f,a,b). Moreover, Coherence Factor (CF), as an effective weighting method for linear-array imaging, was improved in the terms of sidelobes and resolution (Mozaffarzadeh et al., 2017g,h). In the previous publications of the authors, a novel beamforming algorithm based on the expansion of DMAS algebra, namely Double Stage DMAS (DS-DMAS) was introduced, and applied on Photoacoustic imaging (Mozaffarzadeh et al., 2017e,c). In this paper, the performance of DS-DMAS algorithm is evaluated in US imaging. It is shown that expanding DMAS algebra results in summation of the multiple terms which can be treated as a DAS. It is proposed to use DMAS algorithm instead of the existing DAS inside the expansion. The results show that DS-DMAS outperforms DAS in the linear-array US imaging, especially at the presence of high level of the imaging noise and the off-axis signals.

**Methods**

When US signals are detected by linear-array of the US transducers, DAS can be used to reconstruct the image from the detected US signals as follows:

$$y_{DAS}(k) = \sum_{i=1}^{M} x_i(k - \Delta_i), \quad (1)$$

where $y_{DAS}(k)$ is the output of the beamformer, $k$ is the time index, $M$ is the number of the array elements, and $x_i(k)$ and $\Delta_i$ are the detected signals and the corresponding time delay for detector $i$, respectively (Deylami and Asl, 2016). Consequently, DAS reconstructed images contain high level of sidelobes and a low resolution. DMAS was introduced to address the weaknesses of DAS (Matrone et al., 2015). Similar to DAS, DMAS realigns the received RF signals by applying time delays, but the samples are multiplied before adding them up. The DMAS formula is given by:

$$y_{DMAS}(k) = \sum_{i=1}^{M-1} \sum_{j=i+1}^{M} x_i(k - \Delta_i) x_j(k - \Delta_j), \quad (2)$$

To overcome the dimensionally squared problem in (2), the following modifications were suggested (Matrone et al., 2015):

$$\chi_{ij}(k) = \text{sign}[x_i(k - \Delta_i) x_j(k - \Delta_j)] \sqrt{|x_i(k - \Delta_i) x_j(k - \Delta_j)|},$$
$$\text{for} \quad 1 \leqslant i \leqslant j \leqslant M. \quad (3)$$

$$y_{DMAS}(k) = \sum_{i=1}^{M-1} \sum_{j=i+1}^{M} \chi_{ij}(k). \quad (4)$$

Performing sign, absolute and square root after the coupling procedure in (3) and (4), which requires $(M^2 - M)/2$ computations for each pixel, result in a slow imaging. Sometimes these library functions require many clock cycles, causing an improper timing performance of DMAS algorithm. Applying the following procedure to the received US signals reduces the computational load of the sign, absolute and square root operations to $M$ for each pixel (Park et al., 2016):

$$x'_i(k) = \text{sign}[x_i(k)]\sqrt{x_i(k)}, \qquad \text{for} \qquad 1 \leqslant i \leqslant M. \tag{5}$$

$$\chi_{ij}(k) = x'_i(k)x'_j(k), \qquad \text{for} \qquad 1 \leqslant i \leqslant j \leqslant M. \tag{6}$$

A product in the time domain is equivalent to the convolution in the frequency domain. Consequently, new components which centered at the zero and harmonic frequencies are appeared in the spectrum due to the similar range of frequencies for $x_i(k - \Delta_i)$ and $x_j(k - \Delta_j)$. A band-pass filter is applied to the beamformed output signal to attenuate the DC and higher frequency components. It should be noticed that the algorithm is finally called Filtered-DMAS after applying the band-pass filter (Matrone et al., 2015). The output of DMAS beamformer is the spatial coherence of the detected US signals, and the procedure of DMAS algorithm can be considered as a correlation process which uses the auto-correlation of the aperture. In this paper, it is proposed to use DMAS beamforming instead of the existing DAS algorithm inside DMAS algebra. First, consider the expansion of

DMAS algorithm, which can be written as follows:

$$\begin{aligned}
y_{DMAS}(k) &= \sum_{i=1}^{M-1} \sum_{j=i+1}^{M} x_{id}(k) x_{jd}(k) \\
&= x_{1d}(k) \Big[ x_{2d}(k) + x_{3d}(k) + x_{4d}(k) + ... + x_{Md}(k) \Big] \\
&+ x_{2d}(k) \Big[ x_{3d}(k) + x_{4d}(k) + ... + x_{Md}(k) \Big] \\
&+ ... \\
&+ x_{(M-2)d}(k) \Big[ x_{(M-1)d}(k) + x_{Md}(k) \Big] \\
&+ \Big[ x_{(M-1)d}(k) x_{Md}(k) \Big],
\end{aligned} \quad (7)$$

where $x_{id}(k)$ and $x_{jd}(k)$ are the delayed detected signals for the $i^{th}$ and $j^{th}$ elements, respectively. According to (7), there is a DAS in each term of the expansion, which can be used to generate DS-DMAS beamformer as follows:

$$\begin{aligned}
y_{DMAS}(k) &= \sum_{i=1}^{M-1} \sum_{j=i+1}^{M} x_{id}(k) x_{jd}(k) \\
&= \underbrace{\Big[ x_{1d}(k) x_{2d}(k) + x_{1d}(k) x_{3d}(k) + ... + x_{1d}(k) x_{Md}(k)) \Big]}_{\text{the first term}} \\
&+ \underbrace{\Big[ x_{2d}(k) x_{3d}(k) + x_{2d}(k) x_{4d}(k) + ... + x_{2d}(k) x_{Md}(k) \Big]}_{\text{the second term}} \\
&+ ... \\
&+ \underbrace{\Big[ x_{(M-2)d}(k) x_{(M-1)d}(k) + x_{(M-2)d}(k) x_{Md}(k) \Big]}_{\text{(M-2)th term}} \\
&+ \underbrace{\Big[ x_{(M-1)d}(k) x_{Md}(k) \Big]}_{\text{(M-1)th term}}.
\end{aligned} \quad (8)$$

DMAS algorithm is a correlation process in which for each voxel of the image, the calculated delays for each element of the array are combinatorially coupled and multiplied. In other words, the similarity of the samples are obtained. In (8), between all the terms, there is a summation interpreted as the DAS algebra . If the presence of the off-axis signals results in a high error in the correlation process of DMAS, the summation of the calculated correlations leads to a summation of high range of errors. It is proposed to use DMAS beamformer between each term of the expansion instead of DAS. To put it more simply, the samples go through another correlation procedure. To illustrate this, consider (9):

$$y_{DS-DMAS}(k) = \sum_{i=1}^{M-2} \sum_{j=i+1}^{M-1} x_{it}(k) x_{jt}(k), \tag{9}$$

where $x_{it}$ and $x_{jt}$ are the $i^{th}$ and $j^{th}$ terms shown in (8), respectively. The expansion of DS-DMAS beamformer can be written as:

$$\begin{aligned}
y_{DS-DMAS}(k) &= \sum_{i=1}^{M-2} \sum_{j=i+1}^{M-1} x_{it}(k) x_{jt}(k) \\
&= x_{1t}(k) \Big[ x_{2t}(k) + x_{3t}(k) + x_{4t}(k) + ... + x_{(M-1)t}(k) \Big] \\
&+ x_{2t}(k) \Big[ x_{3t}(k) + x_{4t}(k) + ... + x_{(M-1)t}(k) \Big] \\
&+ ... \\
&+ x_{(M-3)t}(k) \Big[ x_{(M-2)t}(k) + x_{(M-1)t}(k) \Big] \\
&+ \Big[ x_{(M-2)t}(k) x_{(M-1)t}(k) \Big].
\end{aligned} \tag{10}$$

Since DAS is a non-adaptive beamformer and considers all the calculated samples for each element of the array identically, the acquired image by every terms would blur the final reconstructed image. Using (10), the blurring would be prevented, and the effects of the noise in the reconstructed images would be reduced. Of note, the same procedure introduced in (5) and (6), is used in DS-DMAS algorithm to speed up the image formation. Also, the necessary band-pass filter is applied to DS-DMAS algorithm to only pass the necessary components generated after the non-linear operations.

**Results and Performance Assessment**

In this section, numerical and experimental results are presented to evaluate the performance of the proposed algorithm in comparison with DMAS and DAS.

*Numerical Results*

Field II simulator is used to generate the simulated signals (Jensen, 1996; Jensen and Svendsen, 1992). A linear-array transducer having 128 elements and a central frequency of 3 $MHz$ is used. The sampling frequency is 100 $MHz$. The impulse response consists of a two-cycle Hanning-weighted sinusoidal waveform, and the excitation pulse consists of a two-cycle sinusoidal waveform. The speed of sound is considered 1540 $ms^{-1}$. Dynamic transmit and receive focusing are synthesized, then the beamforming algorithms are applied to the recorded US signals. Envelope detection, performed by means of the Hilbert transform, has been used at the end for all the presented images. The obtained lines are normalized and then log-compressed to form the final image.

*Wire Targets Phantom*

Wires are positioned in pairs at the depths of 35 $mm$, 40 $mm$, 45 $mm$, 50 $mm$, 55 $mm$ and 60 $mm$. Also, two wires are positioned at the depths of 32 $mm$ and 63 $mm$. Gaussian noise is added to the detected signals, having a $SNR$ of 50 $dB$. As can be seen in Figure 1, DAS leads to a low resolution image along with the high level of sidelobes degrading the quality of the reconstructed image. On the other hand, DMAS reduces the artifacts and sidelobes caused by DAS and improves the quality of the image. Moreover, the point targets are more distinguishable. As shown in Figure 1(c), DS-DMAS results in lower level of sidelobes and artifacts in comparison with DAS and DMAS. Apart from that, DS-DMAS improves the resolution of the formed image and reduces the width of mainlobe. To compare the beamformers in details, the lateral variations at the two depths of imaging are shown in Figure 2. Clearly, DS-DMAS leads to lower level of sidelobes and lateral valley at the both presented depths. Consider, in particular, the depth of 55 $mm$ where the valleys of the lateral variations for DAS, DMAS and DS-DMAS are about -37 $dB$, -51 $dB$ and -70 $dB$, respectively. In other words, DS-DMAS results in 33 $dB$ and 19 $dB$ improvement, in the term of the level of lateral valley, compared to DAS and DMAS, respectively. In addition, the level of sidelobes at the depth of 35 $mm$ for DAS, DMAS and DS-DMAS are about -80 $dB$, -100 $dB$ and -120 $dB$, respectively. To put it more simply, the level of sidelobes, by DS-DMAS, are reduced for about 40 $dB$ and 20 $dB$ in comparison with DAS and DMAS, respectively. Moreover, the narrower width of mainlobe obtained from DS-DMAS is obvious considering the lateral variations. In order to compare the proposed algorithm in the term of noise reduction, Gaussian noise is added to the detected signals, having a $SNR$ of -10 $dB$. The reconstructed images are shown in Figure 3. Clearly, the effects of the added noise degrade the reconstructed image using DAS, and the quality of the image is degraded. DMAS reduces the effects, but the negative effects of noise still degrade the quality of the formed image, shown in Figure 3(b). DS-DMAS results in a higher level of noise reduction in comparison with DAS and DMAS, as can be seen in Figure 3(c), and the quality of the reconstructed image is improved. To compare the beamformers in more details, lateral variations for all the beamformers, at the two depths of imaging, are presented in Figure 4. As shown in both presented lateral variations, DS-DMAS leads to lower level of sidelobes and more distinguishable point targets in comparison with DAS and DMAS. Consider, in particular, the depth of 55 $mm$ where the sidelobe levels of DAS, DMAS and DS-DMAS are for about -50 $dB$, -70 $dB$ and -90 $dB$, respectively. In the other words, DS-DMAS leads to 40 $dB$ and 20 $dB$ level of sidelobes reduction compared to DAS and DMAS, respectively. Moreover, the valley of the lateral variations of DS-DMAS is reduced for about 15 $dB$ and 30 $dB$ in comparison with DMAS and DAS, respectively. To evaluate the proposed method quantitatively, Signal-to-Noise ratio ($SNR$) and Full-Width-Half-Maximum ($FWHM$) metrics for each beamformer are calculated. The points in the right side of the wire target phantoms (having a $SNR$ of -10 $dB$), are used to measure the $SNR$ using the following formula:

$$SNR = 20\log_{10} P_{signal}/P_{noise}, \qquad (11)$$

where $P_{signal}$ and $P_{noise}$ are the difference of the maximum and minimum intensity of each region, and standard deviation of the region, respectively (Üstüner and Holley, 2003). Table 1 shows the calculated $SNR$s, and Table 2 presents the measured $FWHM$s. As can be seen in Table 1, DS-DMAS outperforms DAS and DMAS. Consider, in particular, the depth of 55 $mm$ where $SNR$ for DAS, DMAS and DS-DAMS is about 42.4 $dB$, 56.5 $dB$ and 70.2 $dB$, respectively. In other words, DS-DMAS improves $SNR$ for about 27.8 $dB$ and 13.7 $dB$ compared to DAS and DMAS, respectively. As shown in Table 2, DS-DMAS leads to lower $FWHM$ in the all depths of imaging. Consider, in particular, the depth of 55 $mm$ where $FWHM$ for DAS, DMAS and DS-DMAS results in 1.0 $mm$, 0.8 $mm$ and 0.6 $mm$, respectively. In other words, DS-DMAS leads to 0.4 $mm$ and 0.2 $mm$ improvement in the term of $FWHM$ compared to DAS and DMAS, respectively.

*Cyst Targets Phantom*

Ten cysts, having 4 $mm$ and 2.5 $mm$ radius, are located in five depths of imaging to evaluate the beamformers under the cyst targets. The reconstructed images are shown in Figure 5. As can be seen, the cyst targets are not well detected in the image generated by DAS. Moreover, the reconstructed image suffers from the effects of the noise. Even though DMAS leads to a higher quality image and more detectable cyst targets, in comparison with DAS, the effects of the added noise are still obvious. On the other hand, DS-DMAS suppresses the effects of the noise further, as was shown in wire targets, and it results in a higher image quality compared to DAS and DMAS. It should be noticed that the speckle pattern using DAS seems more uniform since the resolution and sidelobes provided by DAS are not good enough to separate the speckles as they are. However, the low sidelobes of DS-DMAS make the speckle pattern more distinguished. To compare the re-constructed images quantitatively, Contrast Ratio ($CR$) metric is calculated. The calculated $CR$s, for each beamformer, using cysts having 4 $mm$ radius, are presented in Table 3. $CR$ metric is calculated using following equation:

$$CR = 20\log_{10}\left(\frac{\mu_{cyst}}{\mu_{bck}}\right), \qquad (12)$$

where $\mu_{cyst}$ and $\mu_{bck}$ are the means of image intensity before the log compression inside the yellow and red dotted circle in Figure 5, respectively (Matrone et al., 2015). As can be seen in Table 3, DS-DMAS beamforming algorithm leads to a higher $CR$ for all the depths of imaging in comparison with other beamformers. Consider, in particular, the depth of 50 $mm$ where $CR$ is for about -10.9 $dB$, -28.1 $dB$ and -41.6 $dB$ for DAS, DMAS and DS-DMAS, respectively. To evaluate the proposed method in the term of artifacts and sidelobes suppression, and wire detection at the presence of speckle pattern, another simulation has been conducted using a tumor-like object along with a wire target. The reconstructed images are shown in Figure 6. As can be seen, the proposed method improves the image quality and makes the boundaries of the tumor-like object sharper. Moreover, the small single wire target can be detected in the reconstructed image using DS-DMAS, and is not hidden in the background speckle.

*Experimental Results*

Although the numerical results were promising, the proposed algorithm needs to be evaluated using experimental data. RF data from a heart phantom, and a phantom including cysts and wires are used for further investigation. This data was obtained from the Biomedical Ultrasound Laboratory, University of Michigan. Figure 7 shows the reconstructed images using the RF data obtained from the phantom containing cysts and wires. As can be seen in Figure 7, reconstructed image using DAS contains high level of sidelobes and artifacts. Moreover, the area inside the cysts is not clear enough. In Figure 7(b), the formed image using DMAS is shown, and its contrast is improved. By DMAS, although the artifacts inside the cysts are better reduced, compared to DAS, the reconstructed image is still degraded by the artifacts, especially inside the cysts. DS-DMAS improves the contrast of the formed image, and sidelobes are reduced. Moreover, the area inside the cysts is more clear in comparison with DAS and DMAS. To evaluate the experimental results quantitatively, $CR$s are presented in Table 4, for the two cysts in the reconstructed images in Figure 7. As can be seen, for both cyst targets, DS-DMAS outperforms DAS and DMAS in the term of $CR$. Consider, for example, $CR$s for the cyst located at the depth of 65 $mm$, where DS-DMAS leads to 26.3 $dB$ and 14.4 $dB$ improvement in comparison with DAS and DMAS, respectively. In Figure 8, the lateral variations for the cyst located at the depth of 53 $mm$ are presented. As shown, DS-DMAS algorithm outperforms other concerned beamformers. In another experiment, the proposed method is evaluated using a heart phantom. The reconstructed images using RF data obtained from the heart phantom are shown in Figure 9 where artifacts and the high level of sidelobes degrade the quality of the formed image using DAS. On the other hand, DMAS improves the quality of the reconstructed image by reducing the level of sidelobes. DS-DMAS reduces the negative effects of the sidelobes more than DMAS and provides a higher quality.

*Processing Complexity*

In order to compare the proposed method with other beamformers in the term of the computational burden, the number of operations needed for each algorithm are presented in Table 5. As can be seen, the order of the computational complexity for DS-DMAS and DMAS is $O(M^2)$ which is exponentially more than $O(M)$ for DAS. Of note, the overhead computational cost of DS-DMAS would linearly increased by increasing the number of employed elements of the transducer.

**Discussion**

The main improvements obtained by the proposed algorithm are noise reduction and artifacts suppression. The most commonly used beamforming algorithm in US imaging is DAS. Although it provides real time imaging, it results in low quality images. DAS algorithm leads to high level of sidelobes and low resolution images due to its non-adaptiveness and blindness. To put it more simply, DAS considers all the calculated samples for each elements of the array the same as each other. DMAS beamformer, outperforming DAS, was introduced in which the main enhancement was contrast improvement. It improves the reconstructed image due to its correlation process. In fact, DMAS does not consider all the calculated samples for each point of the reconstructed image the same. There is a weighting procedure in DMAS which makes it different from DAS, and that is the reason why it outperforms DAS. Comparing Figure 1(a) and Figure 1(b) clearly demonstrates the improvements gained by DMAS. At the presence of high levels of the imaging noise, DAS beamformer causes a low quality image, suffering from the neg-ative effects of the noise. DMAS reduces the negative effects of noise and improves the image quality due to its correlation procedure which makes this beamforming algorithm non-blind. In other words, each calculated samples is weighted based on other calculated samples, and the final formed image is enhanced. However, it can be perceived that the level of sidelobes and artifacts in DMAS are not satisfying yet, especially at the presence of high level of noise, as can be seen in Figure 3(b) and Figure 5(b). Indeed, the correlation procedure of DMAS is not able to suppress the high level of the off-axis signals and noise well enough. By expanding DMAS formula, it can be seen that DMAS can be written as the multiple summations of various terms. This procedure of summation can be considered as a DAS algorithm. To put it more simply, DAS algebra exists in the expansion of DMAS formula, which can be the source of the low noise and artifacts suppression of DMAS beamformer. Since DMAS outperforms DAS in the terms of level of sidelobes and artifacts, it is proposed to use DMAS instead of the existing DAS algebra inside the expansion of DMAS. DS-DMAS leads to lower level of sidelobes and artifacts compared to DAS and DMAS, especially at the presence of high level of the noise, as a result of double correlation procedure. In fact, in DS-DMAS beamforming algorithm there are two procedures of weighting. At the presence of high level of the imaging noise and artifacts, the first weighting procedure is not able to suppress the negative effects in the formed images, but using two stages of the correlation process, the effects are well suppressed. Figure 2 and Figure 4 show the lateral variations of the beamformers, and clearly DS-DMAS results in lower level of sidelobes and lateral valley, and narrower width of mainlobe in comparison with DAS and DMAS. Table 1, Table 2 and Table 3 show the quantitative comparison of the beamformers, which illustrate the superiority of DS-DMAS using merits of $SNR$, $FWHM$ and $CR$. The proposed method has been evaluated using experimental data. Clearly, DS-DMAS satisfies the expectations and results in the images having a higher contrast and lower levels of noise and artifacts. Of note, all the improvements of DS-DMAS are gained at the expense of the higher computational burden. However, DS-DMAS is at the same order of DMAS. We may assume to implement the algorithm on a FPGA device, e.g. on an Altera FPGA of the Stratix IV family (Altera Corp., San Jose, CA, USA). To implement the floating-point multiplication and the square root operations on double-precision operands, by the available library megafunctions, a latency of 5 and 57 clock cycles would be required to generate respectively the multiplication and square root outputs, with a maximum achievable frequency of 255 $MHz$ and 366 $MHz$ respectively, as reported in the floating-point megafunctions datasheet.

Beamforming plays a significant rule in diagnostic US imaging. In applications in which phased (or micro-convex) arrays are used, since a limited f-number would be available, DS-DMAS can provide a further improvement compared to DMAS. In small-parts and vascular US imaging, for instance *in vivo* imaging of the carotid artery, where the resolution and specially sidelobes are of importance, DS-DMAS can be used, providing higher contrast and noise suppression compared to DAS and DMAS.

**Conclusions**

DAS beamformer is the most common beamfoaming algorithm in US imaging. Even though it provides a real time imaging, it is a blind beamformer and results in low quality images. DMAS beamforming algorithm leads to lower level of sidelobes and a higher resolution in comparison with DAS. However, the off-axis signals and noise still degrade the reconstructed images by DMAS. In this paper, a novel beamforming algorithm based on the expansion of DMAS formula was introduced. It was proposed to use DMAS instead of DAS algebra inside the expansion of DMAS formula. The proposed method has been evaluated numerically and experimentally. The results showed that DS-DMAS leads to lower level of sidelobes of about 25% compared to DMAS. Moreover, the proposed method has been quantitatively evaluated, and in comparison with DMAS beamformer, it was shown that DS-DMAS results in 23%, 22% and 43% improvement in the terms of $SNR$, $FWHM$ and $CR$, respectively.

**Figure Captions**

**Figure 1:** Images of the simulated wire targets phantom using a linear-array transducer. (a) DAS, (b) DMAS and (c) DS-DMAS. All images are shown with a dynamic range of 70 $dB$. Noise was added to the detected signals, having a $SNR$ of 50 $dB$.

**Figure 2:** Lateral variations of DAS, DMAS and DS-DMAS at the depths of (a) 35 $mm$ and (b) 55 $mm$. Noise was added to the detected signals, having a $SNR$ of 50 $dB$.

**Figure 3:** Images of the simulated wire targets phantom using a linear-array transducer. (a) DAS, (b) DMAS and (c) DS-DMAS. All images are shown with a dynamic range of 70 $dB$. Noise was added to the detected signals, having a $SNR$ of -10 $dB$.

**Figure 4:** Lateral variations of DAS, DMAS and DS-DMAS at the depths of (a) 35 $mm$ and (b) 55 $mm$. Noise was added to the detected signals, having a $SNR$ of -10 $dB$.

**Figure 5:** Images of the simulated cyst targets phantom using a linear-array transducer. (a) DAS, (b) DMAS and (c) DS-DMAS. All images are shown with a dynamic range of 70 $dB$. Noise was added to the detected signals, having a $SNR$ of 20 $dB$.

**Figure 6:** Images of the simulated phantom (containing a tumor-like object along with a wire target) using a linear-array transducer. (a) DAS, (b) DMAS and (c) DS-DMAS. All images are shown with a dynamic range

of 70 $dB$. Noise was added to the detected signals, having a $SNR$ of 20 $dB$.

**Figure 7:** Images of the experimental RF data detected by a linear-array transducer (obtained with the wire and cyst targets phantom). (a) DAS, (b) DMAS and (c) DS-DMAS. All images are shown with a dynamic range of 70 $dB$.

**Figure 8:** Lateral variations of DAS, DMAS and DS-DMAS for the experimental cyst located at the depth of 53 $mm$ in Figure 7.

**Figure 9:** Images of the experimental RF data detected by a linear-array transducer (obtained with the heart phantom). (a) DAS, (b) DMAS and (c) DS-DMAS. All images are shown with a dynamic range of 70 $dB$.

**Tables**

**Table 1:** $SNR$ ($dB$) values for the simulated wire targets at the different depths of imaging.

| **Depth** (mm) | DAS | DMAS | DS-DMAS |
|:---:|:---:|:---:|:---:|
| 35 | 49.8 | 64.2 | 78.3 |
| 40 | 47.5 | 61.4 | 75.0 |
| 45 | 45.7 | 59.8 | 73.8 |
| 50 | 44.0 | 58.1 | 71.9 |
| 55 | 42.4 | 56.5 | 70.2 |
| 60 | 41.1 | 55.2 | 69.0 |

**Table 2:** $FWHM$ ($mm$) values for the simulated wire targets at the different depths of imaging.

| **Depth** (mm) | DAS | DMAS | DS-DMAS |
|:---:|:---:|:---:|:---:|
| 35 | 0.6 | 0.5 | 0.4 |
| 40 | 0.7 | 0.5 | 0.4 |
| 45 | 0.8 | 0.6 | 0.5 |
| 50 | 0.9 | 0.7 | 0.5 |
| 55 | 1.0 | 0.8 | 0.6 |
| 60 | 1.1 | 0.9 | 0.7 |

**Table 3:** $CR$ ($dB$) values for the simulated cyst targets at the different depths of imaging.

| **Depth (mm)** | DAS | DMAS | DS-DMAS |
|:---:|:---:|:---:|:---:|
| 10 | -26.4 | -42.0 | -58.2 |
| 20 | -26.1 | -41.8 | -58.0 |
| 30 | -19.7 | -34.1 | - 49.1 |
| 40 | -18.2 | -33.3 | -48.7 |
| 50 | -10.9 | -28.1 | -41.6 |

**Table 4:** $CR$ ($dB$) values for the experimental cyst targets at the different depths of imaging.

| **Depth (mm)** | DAS | DMAS | DS-DMAS |
|:---:|:---:|:---:|:---:|
| 53 | -16.2 | -27.9 | -41.4 |
| 65 | -13.9 | -25.8 | -40.3 |

**Table 5:** Processing complexity for the different beamformers.

| **Beamformer** | Number of Operations |
|:---:|:---:|
| DAS | $M$ |
| DMAS | $\frac{M(M-1)}{2} + 2(M-1)$ |
| DS-DMAS | $M(M-1) + 3(M-1)$ |

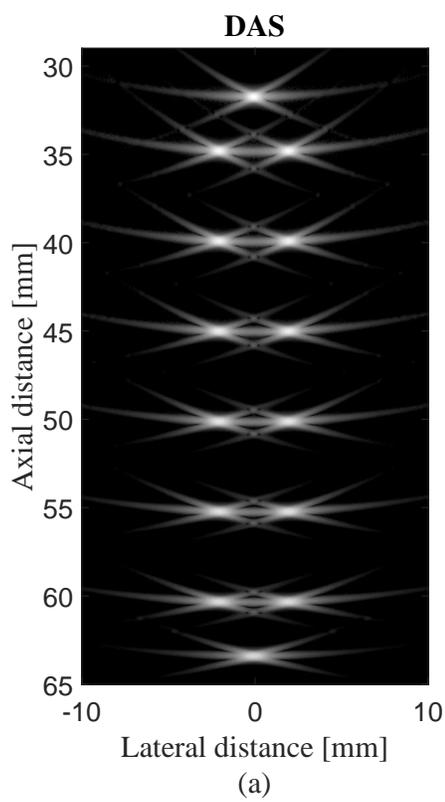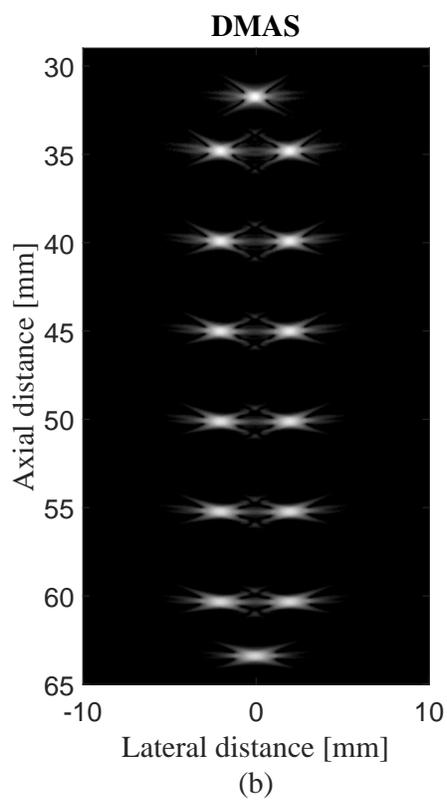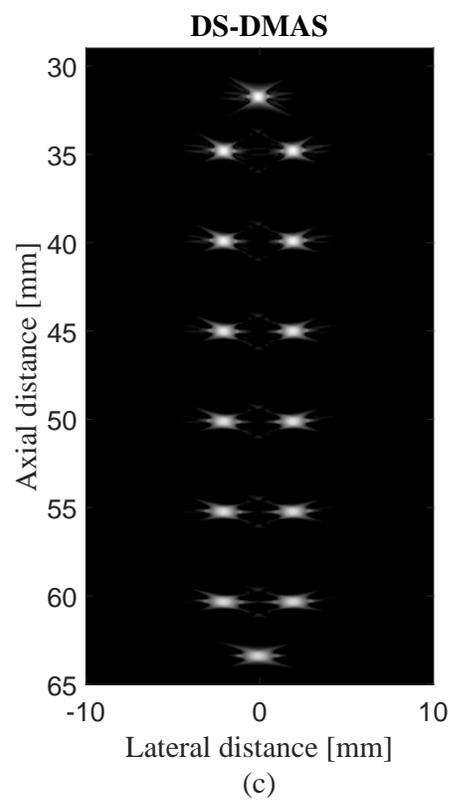

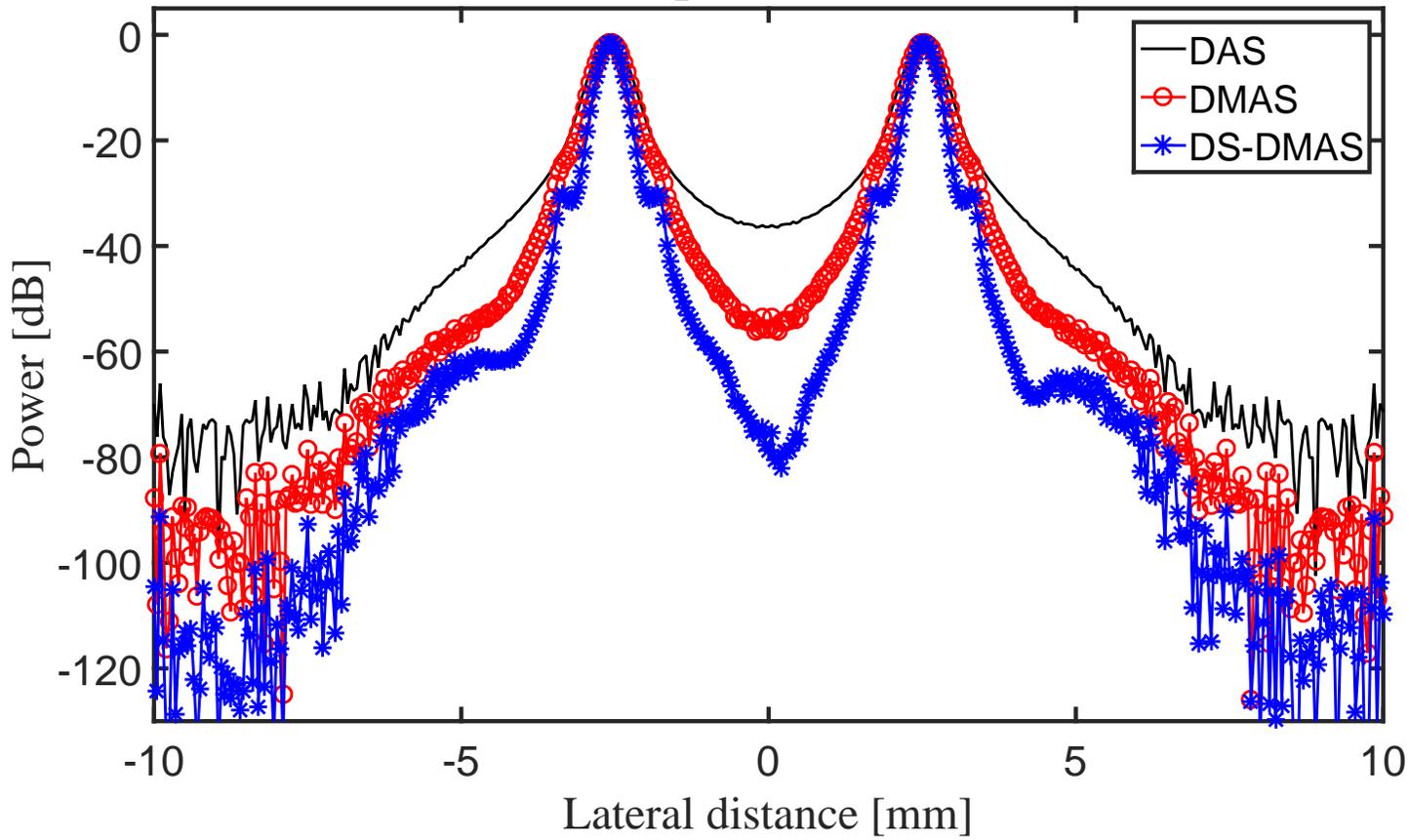

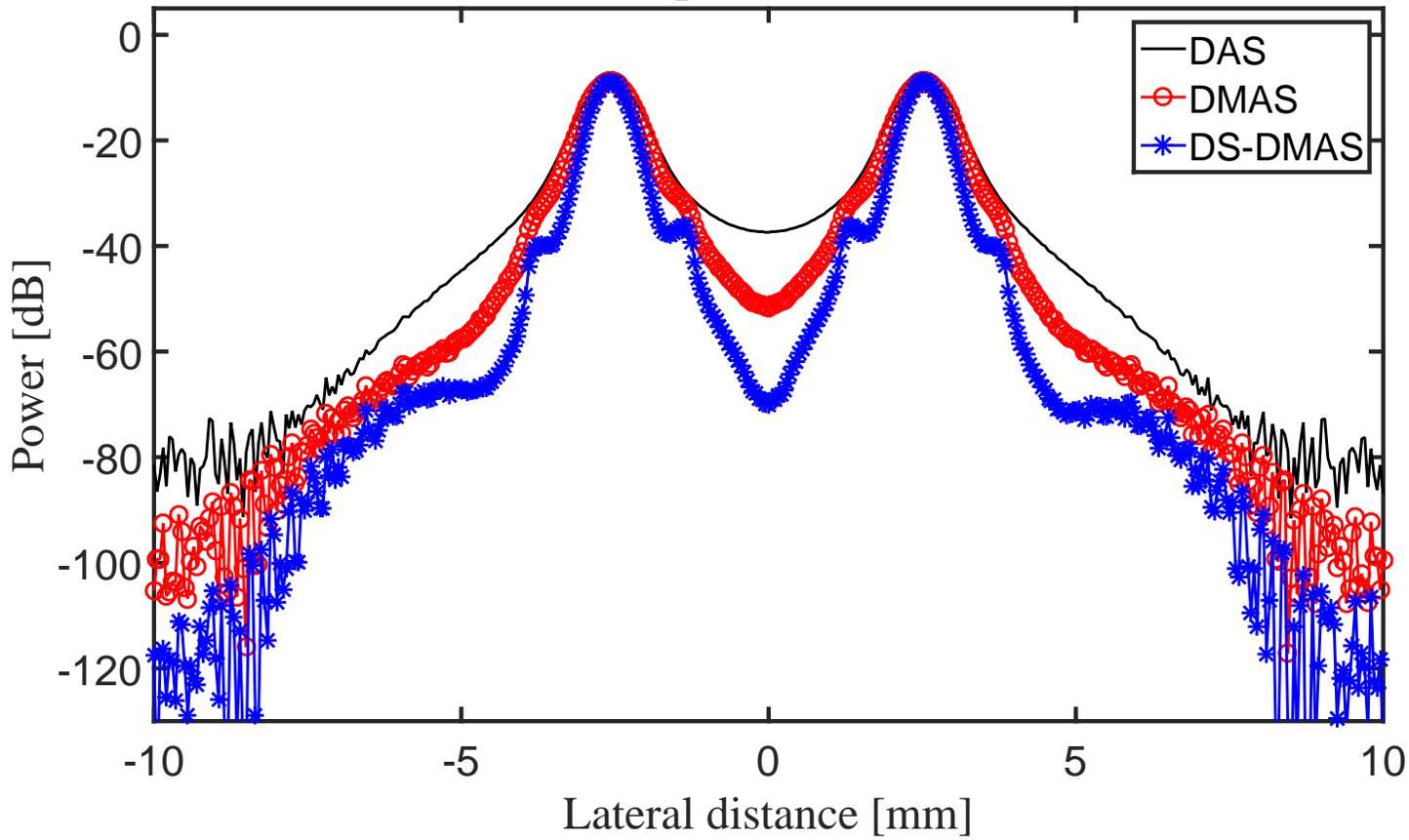

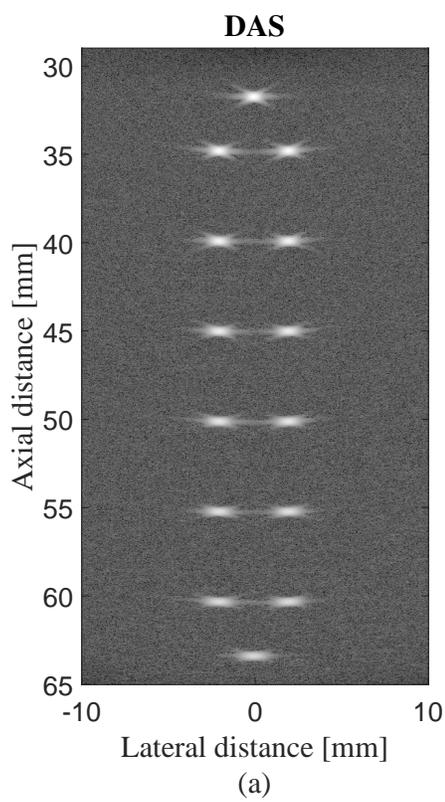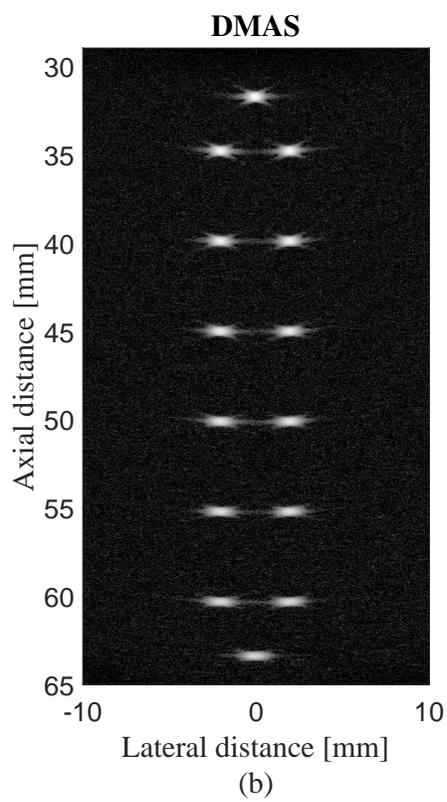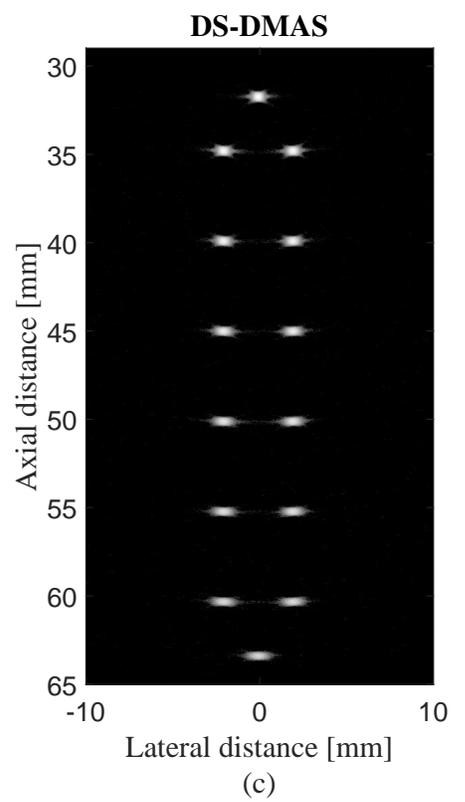

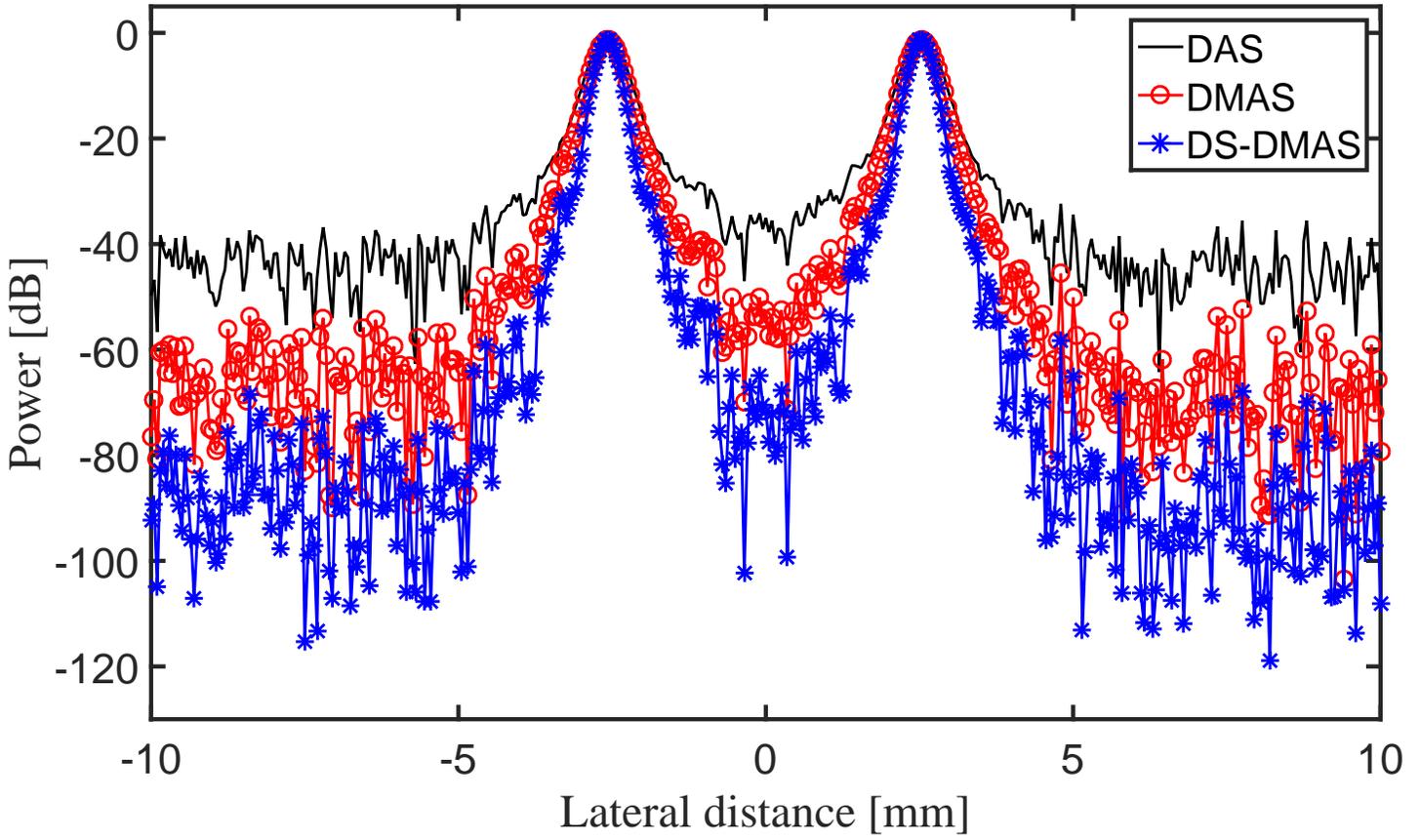

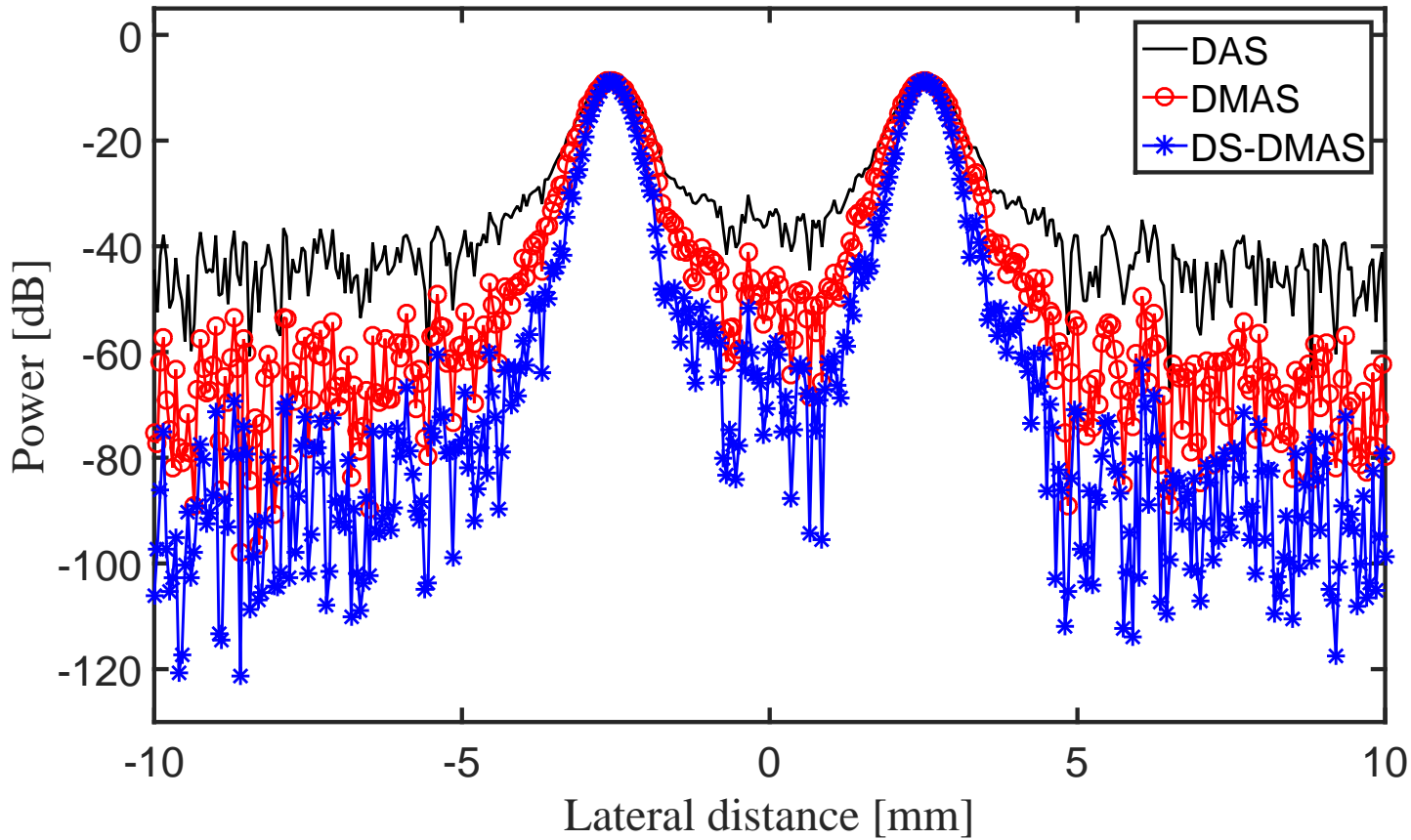

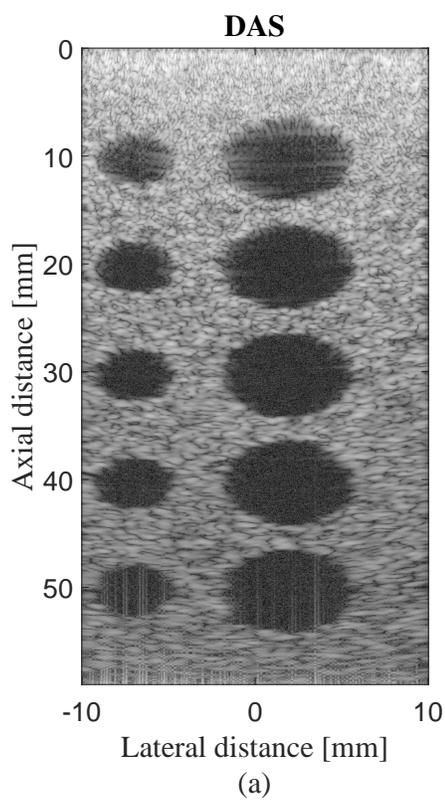 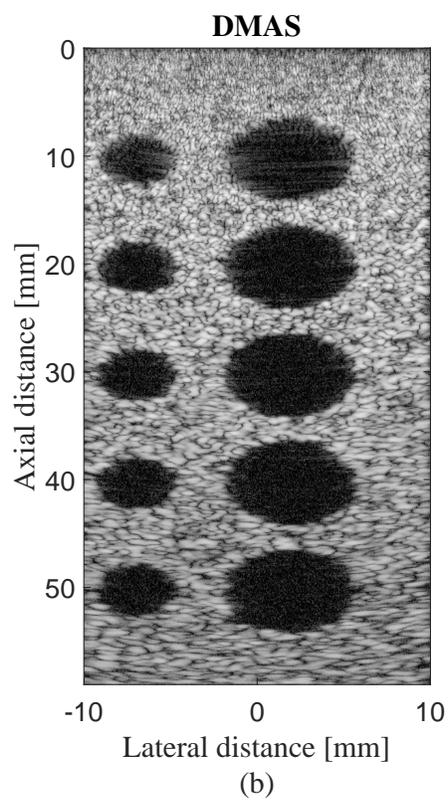 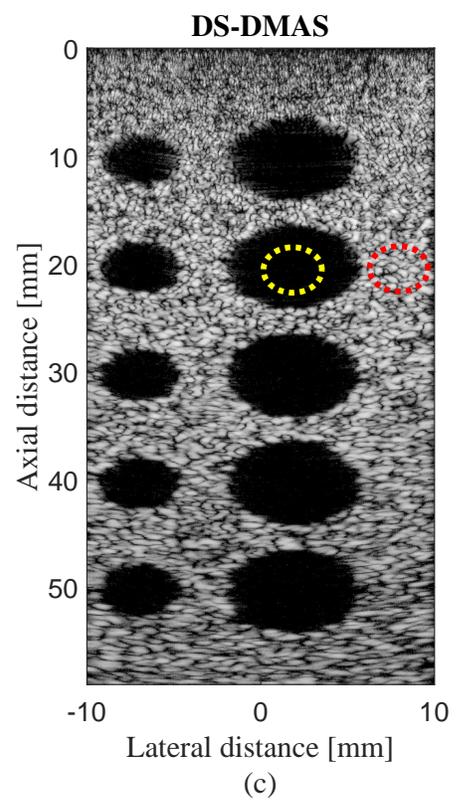

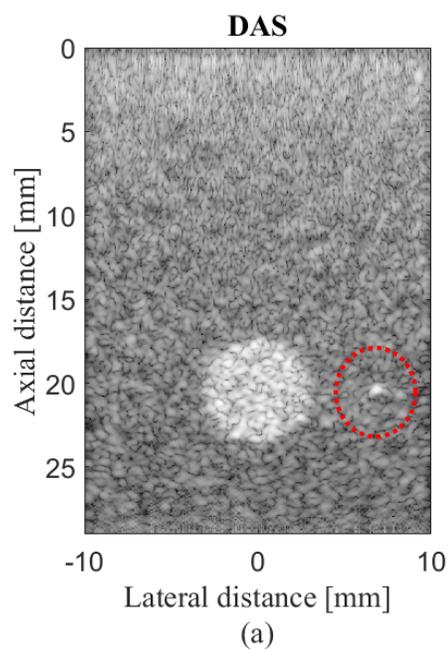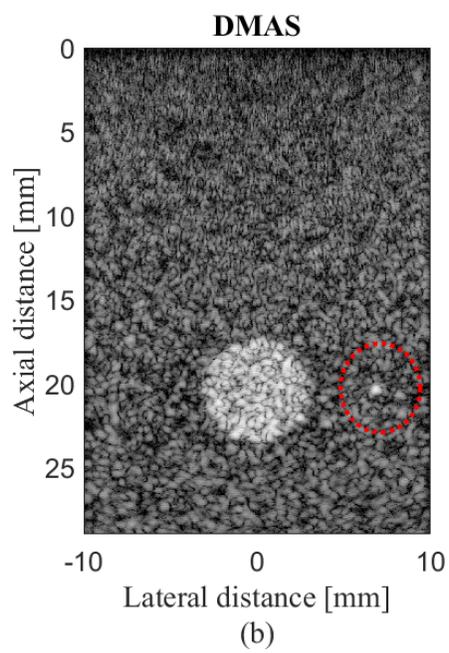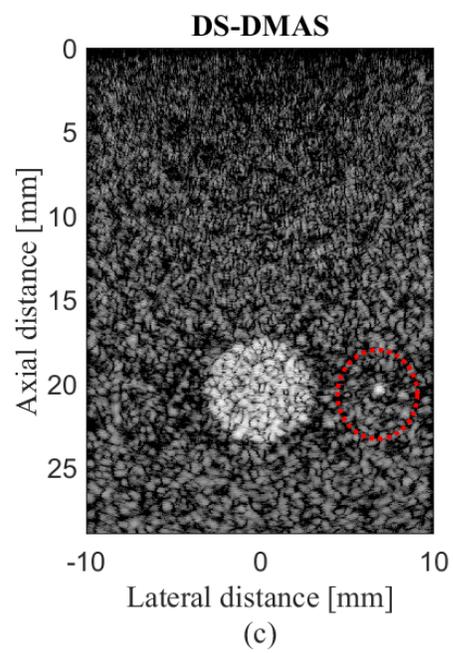

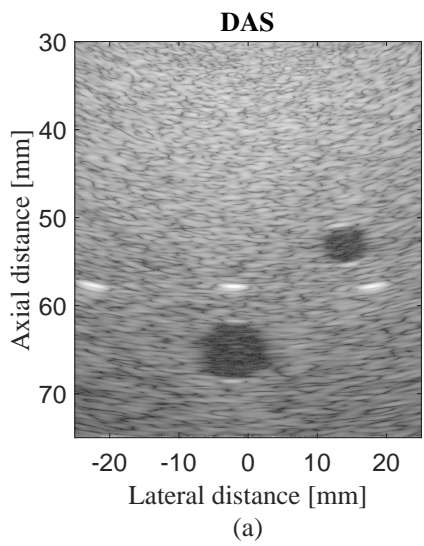 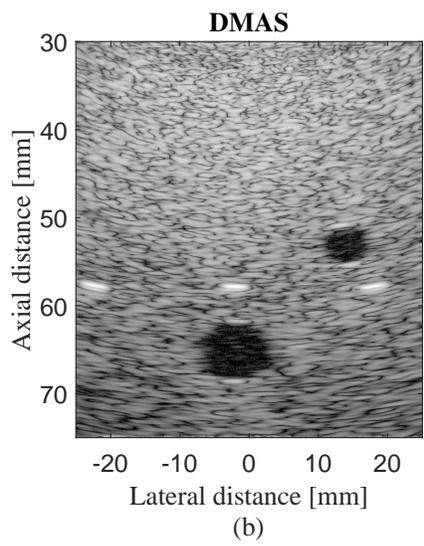 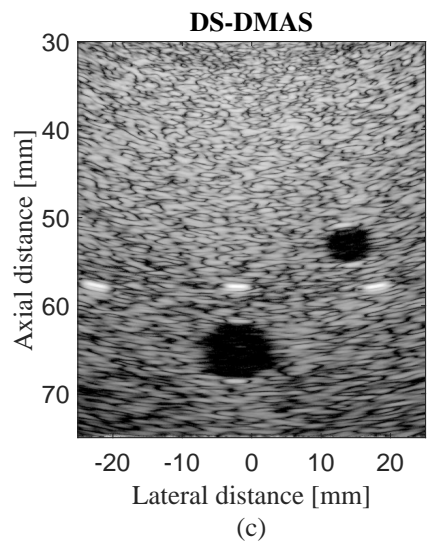

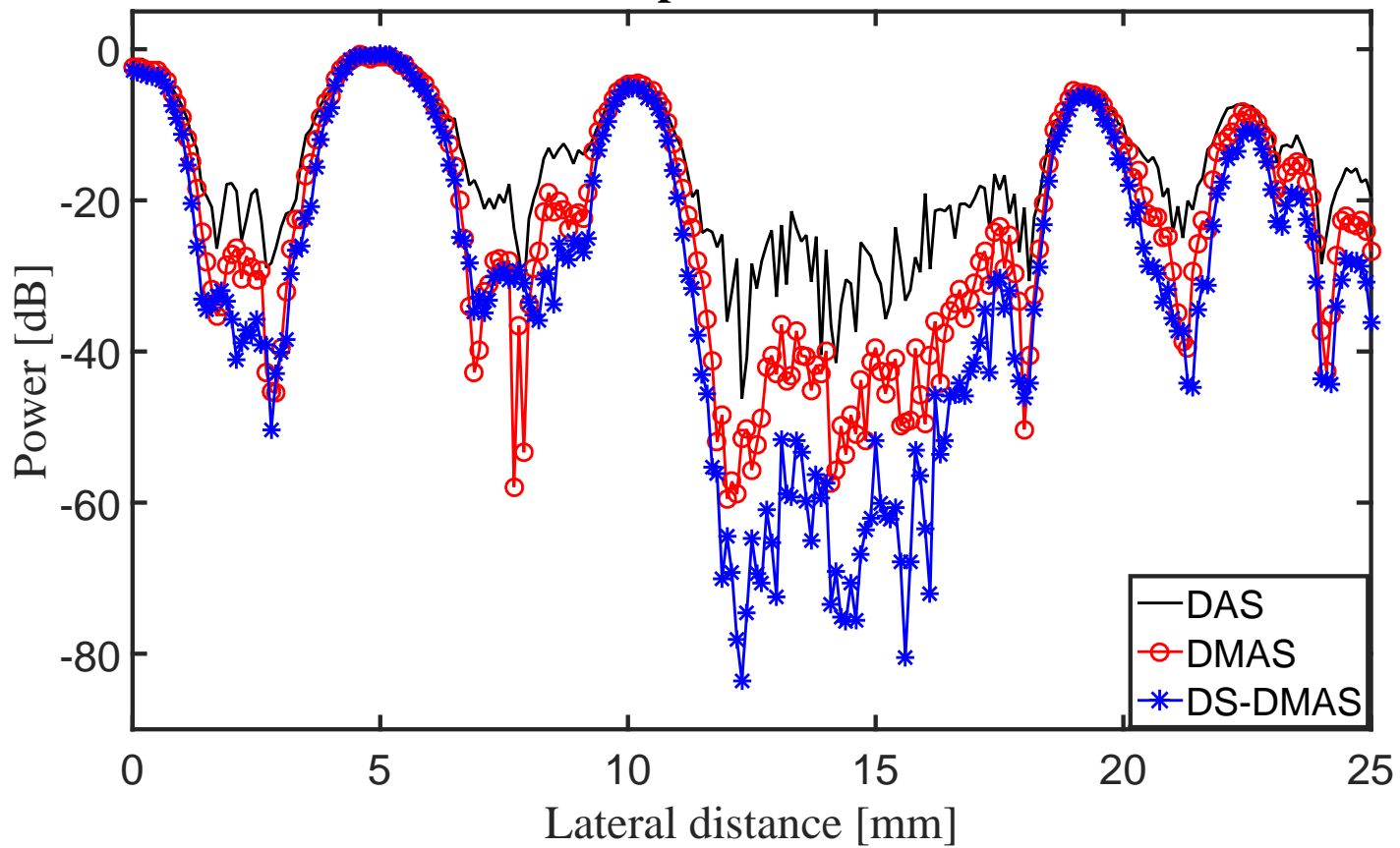

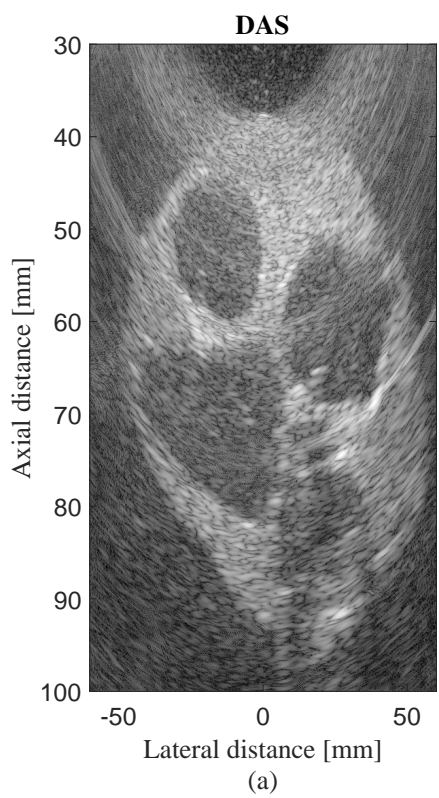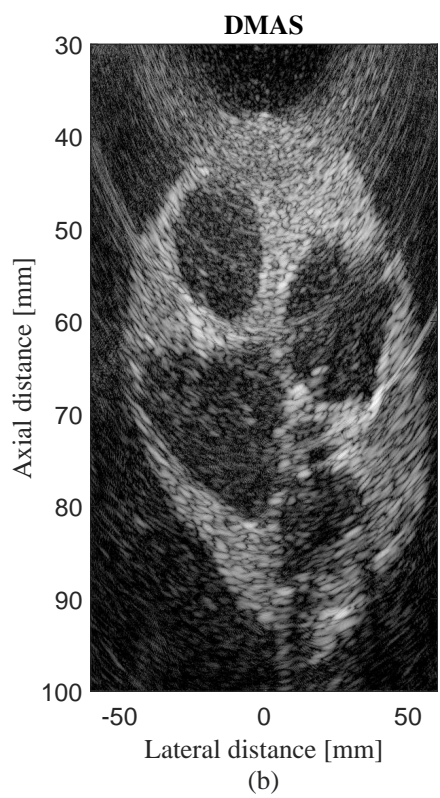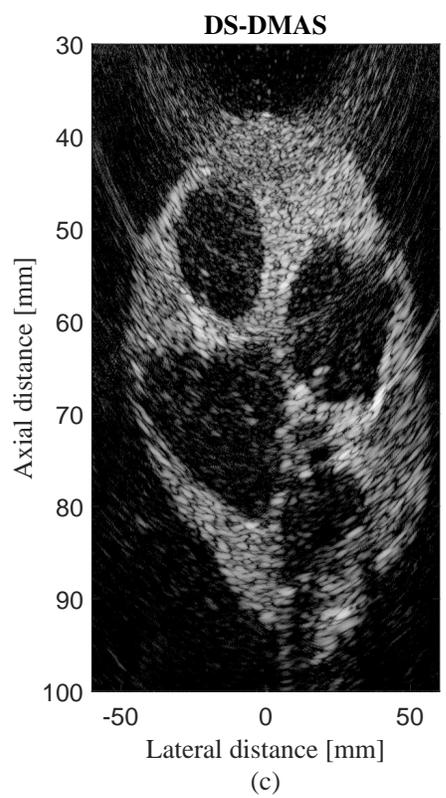